\documentclass[10pt,journal,compsoc]{IEEEtran}
\usepackage{ifpdf}
\ifCLASSOPTIONcompsoc
  \usepackage[nocompress]{cite}
\else
  \usepackage{cite}
\fi
\ifCLASSINFOpdf
  \usepackage[pdftex]{graphicx}
  \graphicspath{{./figs/}}
  \DeclareGraphicsExtensions{.pdf,.jpeg,.jpg,.png}
\else
\fi
\usepackage{amsmath}
\usepackage{gauss}
\newenvironment{sbmatrix}[1]
 {\def\mysubscript{#1}\mathop\bgroup\begin{gmatrix}[b]}
 {\end{gmatrix}\egroup_{\textstyle\mathstrut\mysubscript}}

\usepackage{algorithm}
\usepackage{algpseudocode}
\usepackage{listings}
\usepackage{tcolorbox}

\usepackage{array}
\ifCLASSOPTIONcompsoc
  \usepackage[caption=false,font=footnotesize,labelfont=sf,textfont=sf]{subfig}
\else
  \usepackage[caption=false,font=footnotesize]{subfig}
\fi

\usepackage{amsmath}
\usepackage{algorithm}
\usepackage{color}
\usepackage{comment}
\usepackage{hyperref}
\usepackage{enumerate}
\usepackage{paralist}
\usepackage{times}
\usepackage{epsfig}
\usepackage{amssymb}
\usepackage{verbatim} 
\usepackage{url}
\usepackage{xspace}
\usepackage{grffile}
\usepackage{multicol}

\usepackage{url}
\usepackage[colorinlistoftodos]{todonotes}

\usepackage{enumerate}

\usepackage{rotating}
\usepackage{multirow}
\usepackage{tikz}

\usepackage{comment}

\usepackage{color}

\usepackage{scrextend}

\usepackage{amsfonts}

\begin{document}
\sloppy 
%\title{Design Space Exploration of \\ Oblivious Post-Quantum Storage Systems}
\title{Connecting the Dots: Privacy Leakage via Write-Access Patterns to the Main Memory}


\author{
Tara~Merin~John$^\dagger$,
Syed~Kamran~Haider$^\dagger$,
~Hamza~Omar,
and~Marten~van~Dijk% <-this % stops a space
\IEEEcompsocitemizethanks{\IEEEcompsocthanksitem T. M. John, S. K. Haider, H. Omar, and M. van Dijk are with the Department
of Electrical and Computer Engineering, University of Connecticut, Storrs,
CT, 06279. 
E-mail: \{tara.john, syed.haider, hamza.omar, marten.van\_dijk\}@uconn.edu
\IEEEcompsocthanksitem $\dagger$ T. M. John and S. K. Haider  contributed equally to this work.
\IEEEcompsocthanksitem This work is an extension of an abstract submission, accepted as a poster in HOST 2017, which received ``the best poster award''.
%
%\IEEEcompsocthanksitem D. M. Shila is with United Technologies Research Center, East Hartford, CT, 06118. 
%E-mail: manikad@utrc.utc.com% <-this % stops an unwanted space
}
}

\markboth{}%
{John \MakeLowercase{\textit{et al.}}: Connecting the Dots: Privacy Leakage via Write-Access Patterns to the Main Memory}



\IEEEtitleabstractindextext{%
\begin{abstract}

Data-dependent access patterns of an application to an untrusted storage system are notorious for leaking sensitive information about the user's data.
Previous research has shown how an adversary capable of monitoring both read \textit{and} write requests issued to the memory can correlate them with the application to learn its sensitive data.
However, information leakage through \textit{only} the write access patterns is less obvious and not well studied in the current literature.
In this work, we demonstrate an actual attack on power-side-channel resistant Montgomery's ladder based modular exponentiation algorithm commonly used in public key cryptography.
We infer the complete 512-bit secret exponent in $\mathbf{\sim3.5}$ minutes by virtue of just the write access patterns of the algorithm to the main memory.
In order to learn the victim algorithm's write access patterns under realistic settings, we exploit a compromised DMA device to take frequent snapshots of the application's address space, and then run a simple differential analysis on these snapshots to find the write access sequence.
The attack has been shown on an Intel Core(TM) i7-4790 3.60GHz processor based system. 
We further discuss a possible attack on McEliece public-key cryptosystem that also exploits the write-access patterns to learn the secret key.
%Hence, this work emphasizes a core message that even if only the write access sequence is given, the application's sensitive information can potentially be learned.
%We also discuss some countermeasures to prevent such attacks.

\end{abstract}

}

\maketitle

\IEEEdisplaynontitleabstractindextext

% \fi
\IEEEpeerreviewmaketitle

%*******************************************************%
%          INCLUDE SECTIONS                                                           %
%*******************************************************%
\section{Introduction}	

%first para : explain issues in security, last sentence should allude to side channel attacks.
%second para : describe how side channel attacks work.
%third para  : how such attacks wont be thwarted by current security schemes such as aegis sgx etc
%fourth para : this new DMA side channel, write patterns.
%fifth para : our own attack

%----- Write about general problems of leaking information 

Users' data privacy is becoming a major concern in computation outsourcing in the current cloud computing world.
% with inter-system networks and cloud computing replacing stand-alone local networks.
Numerous secure processor architectures (e.g., XOM~\cite{xom-modelcheck,xom-os}, TPM+TXT~\cite{grawrock-book}, Aegis~\cite{aegis_processor}, Intel-SGX~\cite{intelSGX} etc.) have been proposed for preserving data confidentiality and integrity during a remote secure computation.
The user sends his encrypted data to a secure processor where it is decrypted and computed upon in a tamper-proof environment, and finally the encrypted results of the computation are sent back to the user.

%Such architectures use multiple security schemes (memory encryption) to ensure that attackers do not disrupt system security. 
%For example, XOM based secure architectures [cite papers] propose two primary secure schemes; OTP (one-time-pad) encryption/decryption (for confidentiality) and Merkle tree (for integrity). 
While the secure processors provide sufficient levels of security against \emph{direct} attacks, most of these architectures are still vulnerable to \emph{side-channel} attacks.
%\marten{the sentence below does not flow; do we need references to the mentioned architectures?; both XOM and Aegis are vulnerable to control flow leakage right?}
For instance, XOM and Aegis architectures are vulnerable to control flow leakage via  address bus snooping~\cite{xomefficient, xomfast, xomcaches}. 
Similarly, Intel-SGX, being a strong candidate in secure architectures, is vulnerable to side-channel attacks via a compromised OS\cite{sgxAttack}.

%---------------- Talk about different attacks including read/write patterns

%Current research has demonstrated various attacks~\cite{cachetimingattack, remotetiming, powerattack} on systems, and has proposed their respective attacking prototypes. 
%Such attacks use different analyses~\cite{rsabits,predictattack} to infer the secret information from the memory.
%Various studies have been done to show how sensitive information could be leaked by monitoring both read and write accesses issued to the data outside the processor's trusted boundary, e.g main memory, which may or may not be encrypted~\cite{zhuang2004hide}. 
Zhuang \emph{et al.}~\cite{zhuang2004hide} showed that although the data in the main memory of the system can be encrypted, the access patterns to the memory could still leak privacy.
An adversary who is able to monitor both read \emph{and} write accesses made by an application can relate this pattern to infer secret information of the application.
For example, Islam \emph{et al.}~\cite{islam2012access} demonstrated that by observing accesses to an encrypted email repository, an adversary can infer as much as 80\% of the search queries.
This, however, is a very strong adversarial model which, in most cases, requires direct physical access to the memory address bus. 
In cloud computing, for example, this requires the cloud service itself to be untrusted.
The challenging requirements posed by the above mentioned strong adversarial model leads one to think that applications are vulnerable to privacy leakage via memory access patterns only if such a strong adversary exists, i.e., one capable of monitoring both read \emph{and} write accesses.

%---------------- Talk just about write patterns
In this paper, we counter this notion by demonstrating privacy leakage under a significantly weaker adversarial model.
In particular, we show that an adversary capable of monitoring \emph{only} the write access patterns of an application can still learn a significant amount of its sensitive information.
Hence, in the model of computation outsourcing to a secure processor discussed earlier, even if the cloud service itself is trusted, a remote adversary is still able to steal private information if the underlying hardware does not protect against leakage from write access patterns.

We present a real attack on the famous Montgomery's ladder technique~\cite{joye2002montgomery} commonly used in public key cryptography for modular exponentiation.
Exponentiation algorithms, in general, are vulnerable to various timing and power side-channel attacks~\cite{cachetimingattack, remotetiming, powerattack}.
Montgomery's ladder performs redundant computations as a countermeasure against power side-channel attacks (e.g., simple power analysis~\cite{sparelative}).
However, by monitoring the order of write accesses made by this algorithm, one can still infer the secret \emph{exponent} bits.

%----------------- Propose our own attack
In our weaker adversarial model, since we cannot directly monitor the memory address bus, we learn the pattern of write accesses by taking frequent memory snapshots.
For this purpose, we exploit a compromised Direct Memory Access device (DMA\footnote{DMA grants full access of the main memory to certain peripheral buses, e.g. FireWire, Thunderbolt etc.}) attached to the victim computer system to read the application's address space in the system memory~\cite{aumaitre2010subverting, maynor2005dma, bock2009firewire}. %, maartmann2011inception, boileau2006hit, panholzer2008physical}.
Clearly, any two memory snapshots only differ in the locations where the data has been modified in the latter snapshot.
In other words, comparing the memory snapshots not only reveals the fact that write accesses (if any) have been made to the memory, but it also reveals the exact locations of the accesses which leads to a precise access pattern of memory writes.

% ---------------- Experimental Setup Details
Our experimental setup uses a PCI Express to USB 3.0 adapter attached to the victim system, alongside an open source application called PCILeech~\cite{pcileech}, as the compromised DMA device.
We implement the Montgomery's ladder for exponentiation of a 128 byte message with a 64 byte (512 bits) secret exponent~\cite{rsabits}.
Through our attack methodology, we are able to infer all 512 secret bits of the exponent in just 3 minutes and 34 seconds on average.

Although our experimental setup utilizes a wired connection to a USB 3.0 port on the victim system for DMA, Stewin \emph{et al.} demonstrated that DMA attacks can also be launched \textit{remotely} by injecting malware to the dedicated hardware devices, such as graphic processors and network interface cards, attached to the host platform~\cite{DMAmalware}.
Therefore, our attack methodology allows even remote adversaries to exploit the coarse grained side-channel information obtained by memory snapshots to infer the secret data.
Hence, this effort opens up new research avenues to explore efficient countermeasures to prevent privacy leakage under remote secure computation.

% ---------------- Other examples
%Towards the end of the paper, we further discuss some more examples of applications which could potentially be vulnerable of privacy leakage over write access patterns, e.g., linked lists and binary search trees.

%This paper is aimed to give different attacking strategies for a diverse set of commonly used applications, to leak information via write access patterns. %Moreover, in this paper, we demonstrate an attack on the victim system's memory that leaks private information by monitoring just the write access patterns of Montgomery powering ladder. 
%We take memory snapshots for each and every write access via a compromised Direct Memory Access device (DMA). 
%This could be achieved when the adversary has physical access to the victim machine and can launch an attack by attaching a malicious device capable of DMA\cite{aumaitre2010subverting, boileau2006hit, maynor2005dma} to a running victim system. 
%To monitor these access patterns, we acquire multiple memory snapshots using a PCI express module alongside an open source application~\cite{pcileech}. Towards the end, we present the results of our proposed attack and show how the secret data bits are inferred. 

%However, since this algorithm works with branches driven by secret key bits, write access patterns can be crucial in disclosing secret keys.

\section{Background}\label{sec:background}

%%%%%%%%%%%%%%%%%%%%%%%%%%%%%%%%%%%%%%%%%%%%%%%%%%%%%%%

\subsection{Exponentiation Algorithms}
\begin{algorithm}[!t]
\caption{RSA - Left-to-Right Binary Algorithm} \label{alg:rsa1}
\textbf{Inputs:} $g$, $k=(k_{t-1}, \cdots , k_0)_2$ ~~
\textbf{Output:} $y=g^k$ \\
\textbf{Start:}
\begin{algorithmic} [1]
	\State $R_0 \gets 1$; $R_1 \gets g$ 
	\For{$j=t-1$ \textit{downto} $0$}
		\State $R_0 \gets (R_0)^2$
		\State \textbf{if} $k_j = 1$ \textbf{then} $R_0 \gets R_0 R_1$ \textbf{end if} \label{algoline:rsacondition}
 	\EndFor
\end{algorithmic} 
\textbf{return} $R_0$
\end{algorithm}

%Cryptosystems use the properties of the underlying cryptographic primitives to support the system's security properties.
Exponentiation algorithms have central importance in cryptography, and are considered to be the back-bone of nearly all the public-key cryptosystems. 
Although numerous exponentiation algorithms have been devised, algorithms for constrained devices are scarcely restricted to the square-and-multiply algorithms. 
RSA algorithm, used in e.g. Diffie-Hellman key agreement, is a commonly used exponentiation algorithm which performs computation of the form $y=g^k \mod n$, where the attacker's goal is to find the secret key $k$. 
The commonly used square-and-multiply implementation of this algorithm is shown in Algorithm~\ref{alg:rsa1}.
For a given input $g$ and a secret key $k$, Algorithm~\ref{alg:rsa1} performs multiplication and squaring operations on two local variables $R_0$ and $R_1$ for each bit of $k$ starting from the most significant bit down to the least significant bit.

%\subsection{Exploiting Exponentiation Algorithms}
Notice that the conditional statement on line~\ref{algoline:rsacondition} of Algorithm~\ref{alg:rsa1} executes based on the value of secret bit $k_j$.
%The conditional statements included in these algorithms can be driven based on some secret data (say the exponent). 
Such conditional branches result in two different power and timing spectra of the system for $k_j=0$ and $k_j=1$, hence leaking the secret key $k$ over the timing/power side-channels.
Similar attacks~\cite{rsabits} have leaked 508 out of 512 bits of an RSA key by using branch prediction analysis (BPA).
%Apart from this, when encrypting with low encryption exponents (e.g., $x$ $=$ 3) and small values of the $y$, $R$ comes out to be strictly less than the modulus $n$. In this case, ciphertexts can be easily decrypted by taking the $x^{th}$ root of the ciphertext over the integers.
Thus, the attack-prone nature of RSA algorithm (Algorithm~\ref{alg:rsa1})
%is prone to many different attacks and can leak information based on the time or power consumption. 
%This produced 
poses a need for an alternate secure algorithm.
% with same functionality however, ensuring security. 

%\begin{algorithm}
% \caption{RSA - Right-to-Left Binary Algorithm}
%\label{alg:rsa2}
% \begin{algorithmic} [1]
% \State \textbf{Input:} $g$, $k$ $=$ $k_{t-1}$,...., $k_0$ 
% \State \textbf{Output:} $y$ $=$ $g^k$ 
% \State $R_0$ $=$ 1; $R_1$ $=$ g;
% \For {$j$ $=$ $0$ \textit{to} $(t-1)$}
%% \For $j$ $=$ $t$ \textit{downto} 1 \textit{do}
% \If {($k_j$ $=$ 1)}
% \State $R_1$ $=$ $(R_1)^2$;
% \EndIf
% \EndFor
% \State \textbf{return} $R_0$
% \end{algorithmic} 
% \end{algorithm}

\subsection{Montgomery's Power Ladder Algorithm}
\begin{algorithm}[!t]
\caption{Montgomery Power Ladder Algorithm} \label{alg:mont}
\textbf{Inputs:} $g$, $k=(k_{t-1}, \cdots , k_0)_2$ ~~
\textbf{Output:} $y=g^k$ \\
\textbf{Start:}
\begin{algorithmic}[1]
	\State $R_0 \gets 1$; $R_1 \gets g$ \label{algostep:declaration}
	\For{$j=t-1$ \textit{downto} $0$}
		\If{$k_j = 0$} ~~ $R_1 \gets R_0 R_1$; ~ $R_0 \gets (R_0)^2$
		\Else ~~~~~~~~~~~~~~~ $R_0 \gets R_0 R_1$; ~ $R_1 \gets (R_1)^2$
		\EndIf
	\EndFor
\end{algorithmic}
\textbf{return} $R_0$
\end{algorithm}
Montgomery Power Ladder \cite{joye2002montgomery} shown in Algorithm~\ref{alg:mont} performs exponentiation without leaking any information over power side-channel. 
Regardless of the value of bit $k_j$, it performs the same number of operations in the same order, hence producing the same power footprint for $k_j = 0$ and $k_j=1$.
Notice, however, that the specific order in which $R_0$ and $R_1$ are updated in time depends upon the value of $k_j$.
E.g., for $k_j = 0$, $R_1$ is written first and then $R_0$ is updated; whereas for $k_j = 1$ the updates are done in the reverse order.
This sequence of write access to $R_0$ and $R_1$ reveals to the adversary the exact bit values of $k$.
%Because of its high regularity and ability to withstand various side-channel attacks (as it performs equal number of multiplications in each branch taken regardless of what the exponent bit is), it is considered to be an elite choice for secure exponentiation. However, 
In this paper, we exploit this vulnerability in a real implementation of Montgomery ladder to learn the secret key $k$.
%demonstrate and present that montogomery power ladder, infact being resistant to side-channel attacks, can leak secret data and information via write access patterns. The implementation of the montgomery ladder is shown in algorithm~\ref{alg:mont}.

%\begin{algorithm}
% \caption{Montogomery Power Ladder Algorithm}
%\label{alg:mont}
% \begin{algorithmic} [1]
% \State \textbf{Input:} $g$, $k$ $=$ $k_{t-1}$,...., $k_0$ 
% \State \textbf{Output:} $y$ $=$ $g^k$ 
% \State $R_0$ $=$ 1; $R_1$ $=$ g;
% \For {$j$ $=$ $(t-1)$ \textit{downto} 0}
% \If {($k_j$ $=$ 0)}
% \State $R_1$ $=$ $R_0R_1$; $R_0$ $=$ $(R_0)^2$;
% \Else
% \State $R_0$ $=$ $R_0R_1$; $R_1$ $=$ $(R_1)^2$; 
% \EndIf
% \EndFor
% \State \textbf{return} $R_0$
% \end{algorithmic} 
% \end{algorithm}

%Montgomery algorithm works in such a way that it not only alters the result to be computed but the base as well gets modified in the process. Moreover, for each branch taken based on the exponent bit, there is a specific order associated to the branch in which the result and base get modified. It is clear from the public algorithm~\ref{alg:mont}, when the exponent bit is 0 the base gets modified first then the result, and otherwise for exponent bit 1. 

\section{The Proposed Attack} \label{sec:attack}

\subsection{Adversarial Model}

\begin{figure}[!t]
%\begin{figure}[h!]
\centering
\includegraphics[width=0.9\columnwidth]{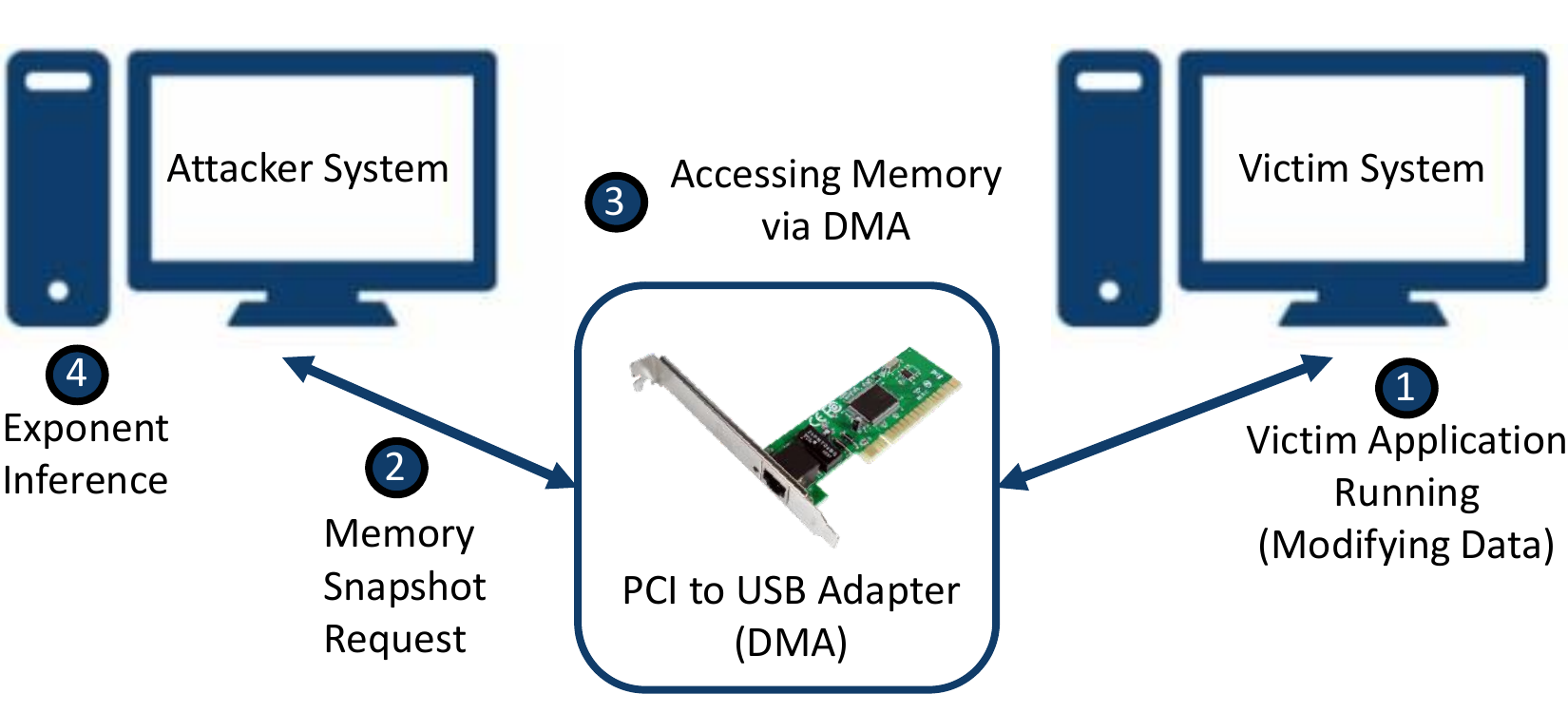}
\caption{Our adversarial model: The attacker system takes snapshots of the victim's DRAM via the PCI adapter to infer the secret key.}
\label{fig:attack}
\end{figure}

Consider a computer system that is continuously computing  exponentiations of the form $y=g_i^k$ for the given inputs $g_i$ using the same secret exponent $k$ according to Algorithm~\ref{alg:mont}.
We call this system the \emph{victim} system.
All the data stored in the main memory of this system is encrypted.
Let there be a compromised DMA device (e.g., a PCI-to-USB adapter) connected to the victim system through which an attacker system can read the whole main memory of the victim as shown in \figurename~\ref{fig:attack}.
The attacker system, however, is limited in its ability to successively read the victim's memory by the data transfer rate of the underlying DMA interface.
The adversary's goal is to find the key $k$ by learning the application's write pattern through frequent snapshots of the victim system's memory.
%For simplicity, we assume that the victim system either has \emph{write-through} caches or caching is disabled altogether, hence any writes by its applications are always propagated to the DRAM.
%\textcolor{blue}{
The victim system used in our attack comes with a \emph{write-through} cache configuration enabled by default.
As a result, any write operations performed by the application are immediately propagated through the memory hierarchy down to the untrusted DRAM.
%has inherently enabled \emph{write-through} cache policy. Due to such characteristics of the system, any modifications made (write operations) by the application, in the memory,  are always propagated to the DRAM.}
Furthermore, we assume that the victim application receives all the inputs $g_i$ in a batch and continuously produces the corresponding cipher texts such that the physical memory region allocated to the application during successive encryptions remains the same.
In other words, the application is not relocated to a different physical address space by the OS throughout the attack.
Such use cases can be found in the applications that require computing signatures of large files.

\subsection{Attack Outline}

Given the above mentioned setting, we proceed with our attack methodology as follows: 
First, a full scan of the victim's memory is performed to identify the physical address space allocated to the victim's application.
Since the adversary requires victim application's memory snapshots at a high frequency, it is infeasible for him to always read the full victim memory because of the data transfer rate being the frequency limiting factor.
Once the address space is identified, the next step is to identify the two memory regions allocated to each of the local variables $R_0$ and $R_1$ (cf. Algorithm~\ref{alg:mont}) within the victim application's address space.
This allows any observed change in either of these two regions to be linked with an update to the variables $R_0$ and $R_1$ respectively.
Finally, the updates in $R_0$ and $R_1$ memory regions are observed via frequent snapshots for a period of one complete encryption, and the order of these updates is linked back to Algorithm~\ref{alg:mont} to learn the key $k$.
We explain these steps in detail in the following subsections.

%Figure~\ref{fig:attack} shows the scenario in which an attacker system could potentially infer victim's secret information. 
%To obtain the secret data, the attacker takes multiple snapshots of the victim's memory while the application is running on the system. 
%One possible way to obtain these snapshots could be via direct memory access (DMA). 
%The \emph{PCI module} shown in the figure is connected to the PCI express slot of the victim giving full access to the physical memory. 
%This module helps in monitoring the victim's memory by taking the snapshots. 
%The victim system runs the montogomery algorithm (shown in algorithm~\ref{alg:mont}) over \emph{x-bit} base (message) and \emph{y-bit} exponent (key) to compute a result of \emph{xy-bits}. 
%In order to compute the result, the application has to traverse over all the y-bits of the exponent one by one. 
%Based on the bit value, it performs computations over the base and the intermediate result. 
%Assuming the physical memory and the data residing in it to be encrypted, the attacker cannot directly compute the exponent via back-tracing along the ciphertexts. 
%However, with the information regarding this multiplication order, and the frequent snapshots of the memory, the write access patterns can reveal much information for an attacker to infer about the exponent or secret data even though the data is encrypted. 

\subsection{Step 1: Application's Address Space Identification} \label{sec:step1}

\begin{algorithm}[!t]
\caption{Victim App's Address Space Identification} \label{alg:first}
\textbf{Inputs:} $M$: Set of memory blocks to scan. \\
\textbf{Output:} $S$: Set of application's memory block(s). \\
\textbf{Start:}
\begin{algorithmic} [1]
\State $S = \emptyset $ \Comment{Initially empty set.}
\For{$m \in M$} \Comment{Scan each block.}
 \State $s_1 = $ \Call{TakeSnapshot}{$m$}
 \State $s_2 = $ \Call{TakeSnapshot}{$m$}
 \If {\Call{CompareMatch}{$s_1, s_2$}}
 	\State $S =  S \cup m$
 \EndIf
 \EndFor
 \State \textbf{return} $S$
 \end{algorithmic} 
 \end{algorithm}

Since the application is supposed to be continuously updating its data (e.g., variables $R_0$, $R_1$), its address space can be identified by finding the memory regions which are continuously being updated.
%the goal is to prune out a major portion of victim's memory which does not get modified. 
Algorithm~\ref{alg:first} shows this process at an abstract level.
The whole of the victim system's memory space is divided into $M$ blocks, each of some reasonable size $B$ (say a few megabytes). 
Two subsequent snapshots of each block $m \in M$ are compared with each other through \Call{CompareMatch}{} procedure.
It is a heuristic based process which searches for a specific pattern of updates between the two snapshots which potentially represents the application's footprint.
For example, a sequence of two modified consecutive $64$ byte cache lines followed by a few unmodified cache lines and then further two modified consecutive cache lines would potentially represent the two $128$ byte regions for $R_0$ (first two cache lines) and $R_1$ (last two cache lines).
Finally, a set $S$ of all those memory blocks which show the specific update sequence searched by \Call{CompareMatch}{} is returned.
This algorithm is iteratively repeated until a reasonably small set of memory block(s) (e.g., one $4$ kB page) is identified which is expected to contain the victim application's address space.

%Critical region of the memory are those where the data is continuously being altered. This phase involves learning such critical/potential regions of the memory to decide which parts of the memory are being accessed regularly. Through this iterative step, we narrow down to one or two regions to monitor from the whole victim's memory. 

\subsection{Step 2: Distinguishing Local Variables $R_0$ and $R_1$} \label{sec:step2}

\begin{figure}[!t]
\centering
\includegraphics[width=0.9\columnwidth]{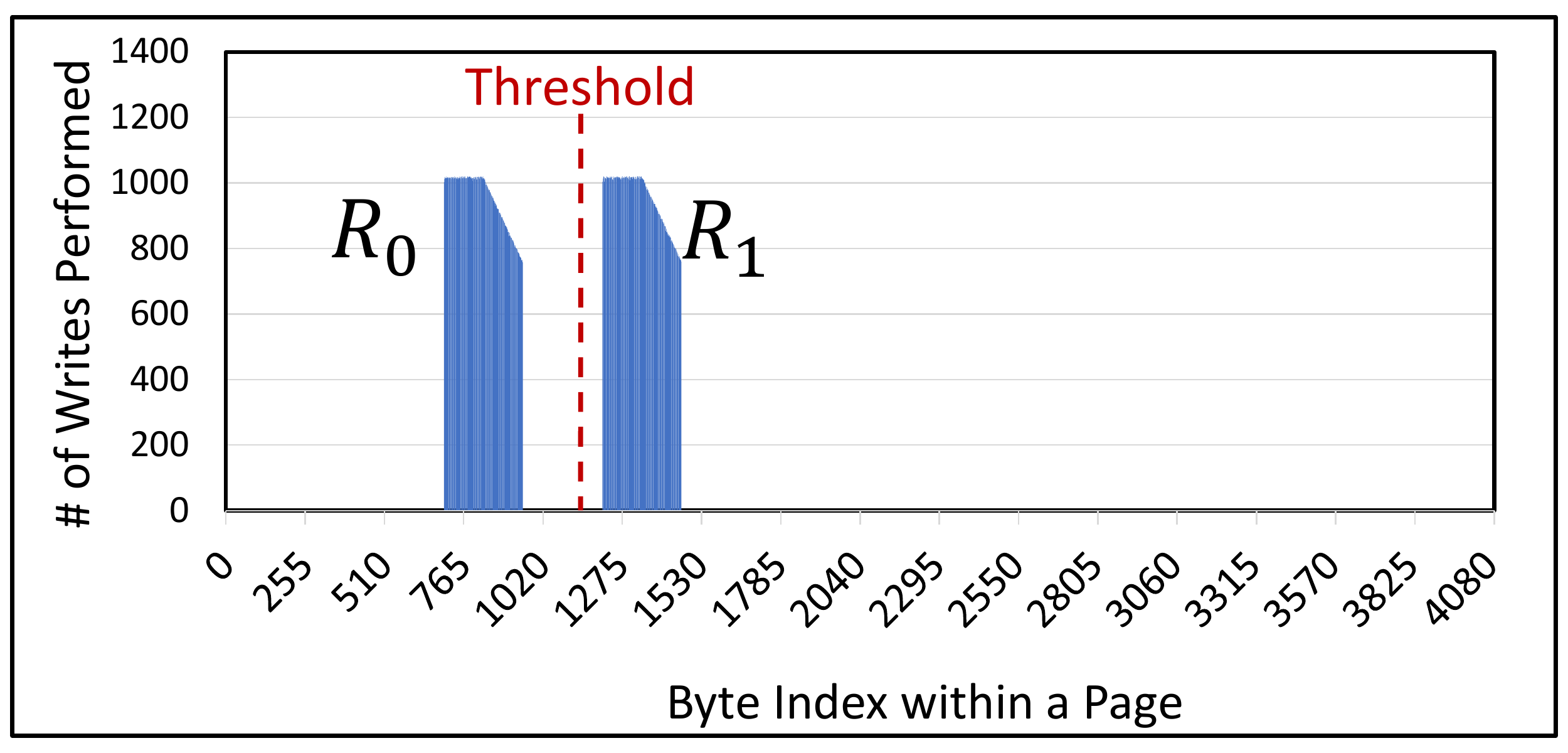}
\caption{A histogram of \# of writes to individual bytes in the victim's memory page. A clear distinction is shown between the regions corresponding to variables $R_0$ and $R_1$.}
\label{fig:histogram}
\end{figure}

Once the application's memory space is found, we need to link two distinct regions within this address space to the variables $R_0$ and $R_1$ in order to determine the key bits from the order of their updates.
For this purpose, a set $V$ of $n$ snapshots of the application's space is computed as shown in Algorithm~\ref{alg:second}. 
Notice that $n$ is large enough to cover one full encryption period.
The \Call{ComputeThreshold}{} procedure computes a histogram of the updates performed inside the application's memory over all the snapshots of set $V$.
\figurename~\ref{fig:histogram} shows one such histogram for a 4kB page of victim's memory. 
It can be seen that almost all the updates are performed at two distinct regions spanning over only a few cache lines within the page. 
These two regions correspond to the variables $R_0$ and $R_1$ respectively\footnote{We can tell whether $R_0$ or $R_1$ comes first in the memory layout from the declaration order of these variables in the actual implementation of the exponentiation algorithm (cf. line~\ref{algostep:declaration} in Algorithm~\ref{alg:mont}).}.
The \emph{inactive} region between $R_0$ and $R_1$ represents a threshold which is later used by \Call{Correlate}{} procedure to determine whether a change in two successive memory snapshots corresponds to an update in $R_0$ or $R_1$ etc.

\begin{algorithm}[!t]
 \caption{Pseudo code for the second phase of attack} \label{alg:second}
 \textbf{Input:} $S$: Application's memory space. (from~Algorithm~\ref{alg:first}); $~~~~~~~~\,~~~~~n$: \# of snapshots to cover one full encryption period. \\
 \textbf{Output:} $k$: Application's secret key. \\
\textbf{Start:}
\begin{algorithmic} [1]
\State $V = (s_1, \cdots, s_n)~|~s_i=\Call{TakeSnapshot}{S}, 1\le i \le n$
\State $Th = $ \Call{ComputeThreshold}{$V$}
\State $W = \emptyset$, $k=(0, \cdots, 0)$ 

\State $V = \Call{RemoveUnchangedSnapshots}{V}$ \label{algostep:inference}
\For {$i = 1 \textbf{ to } |V|-1$} 
 	\State $R_{x_i} = \Call{Correlate}{s_i, s_{i+1}, Th}$ \Comment{$x_i \in\{0,1\}$}
	\State $W = W \cup R_{x_i}$
\EndFor
 
 \State $i=1$, $j=0$
 \For {$(R_{x_i}, R_{x_{i+1}}) \in W$}
 	\If {$R_{x_i} = R_0$ and $R_{x_{i+1}} = R_1$}  $k_j = 1$
	\ElsIf {$R_{x_i} = R_1$ and $R_{x_{i+1}} = R_0$} $k_j = 0$
	\EndIf
	\State $i=i+2$; $~~j=j+1$
\EndFor
\end{algorithmic} 
\textbf{return} $k$
\end{algorithm}

\subsection{Step 3: Inferring the Secret Key} \label{sec:step3}
\begin{figure}[!t]
\centering
\includegraphics[width=1.0\linewidth]{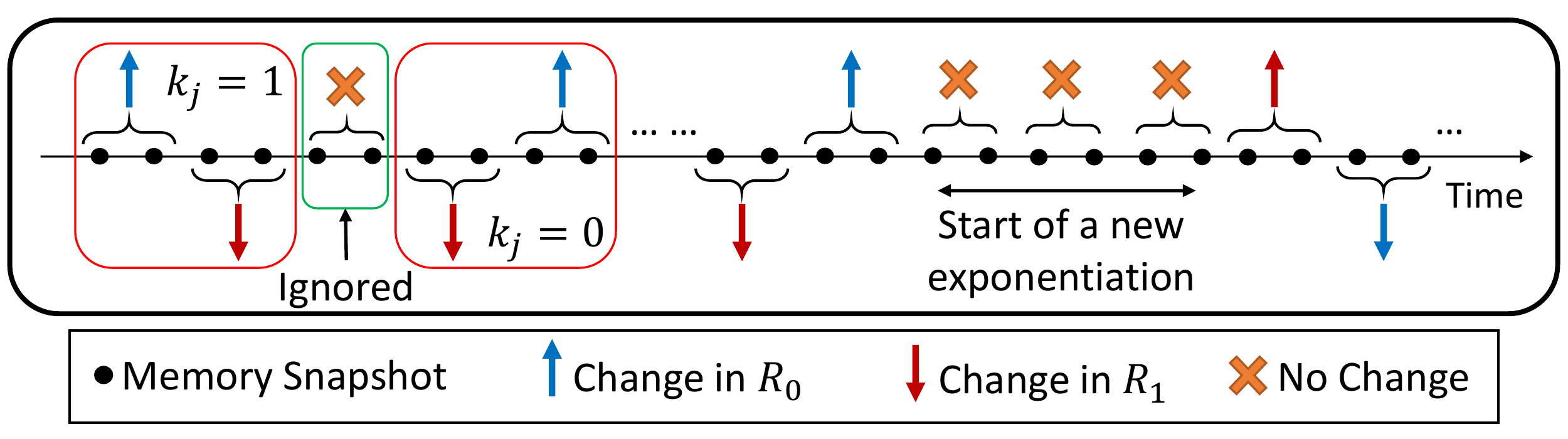}
\caption{Inferring the secret key via observing the sequence of snapshots and the changes in variables $R_0$ and $R_1$. The pairs of snapshots which do not show any change are ignored.}
\label{fig:infer}
\end{figure}

After computing the set of snapshots $V$ and the threshold $Th$, we enter the final phase of inferring the secret key (starting from step~\ref{algostep:inference} in Algorithm~\ref{alg:second}).
Up to this point, the sequence $V$ contains pairs of snapshots that represent changes in $R_0$ and $R_1$, and also the pairs which represent no change, as shown in \figurename~\ref{fig:infer}.
The reason why some pairs do not show any change is because our snapshot frequency is higher than the rate at which the application updates its data.
This allows us to learn the write access pattern at a fine granularity.

In order to learn the write access pattern, first the pairs of unchanged snapshots from the sequence $V$ are removed by the procedure \Call{RemoveUnchangedSnapshots}{$V$}.
The resulting sequence $V$ only contains pairs which always represent a change, either in $R_0$ or $R_1$.
Now, each pair of two successive snapshots is correlated to an update in either $R_0$ or $R_1$ by \Call{Correlate}{} procedure using the threshold computed earlier. 
%\marten{Algorithm 4 is confusing because x can be 0 or 1 in line 5 but the same x is used twice in line 11 where it reflects a 0 and a 1. You can also use $R_{x_i}$ which directly links indices to the variables $x_i$.}

As mentioned earlier, Montgomery ladder algorithm performs computations upon local variables, where the order of variable updates is based on the secret exponent bits (cf. Algorithm \ref{alg:mont}). 
%\marten{This only works if the frequency at which page snapshots are taken neatly fits the way DRAM is updated by the Montgomery algorithm. Is this a result of setting parameters properly or of how a DRAM controller operates (read/write requests are served FIFO or something).}
%Result and base both get altered regardless of which branch we take, only the order changes as shown in algorithm~\ref{alg:mont}. 
Therefore, judging from the order of updates made in $R_0$ and $R_1$, each pair of updates $(R_{x_i}, R_{x_{i+1}}) \in W$ is linked back to the corresponding value of the secret key bit $k_j$ as shown in figure~\ref{fig:infer}. 
%Our inference get more evident as the granularity of the snapshots becomes fine. 
%We could observe much better deduction regarding the inference of exponent bits, if the snapshots are taken at fine granular level.
As the set $W$ contains the history of all the updates to $R_0$ and $R_1$ for a complete encryption, therefore all the key bits can be inferred through the above mentioned process. 

%After determining the critical region(s), we take multiple snapshots of these region(s) and compare them on the fly. 
%The number of differences in two consecutive snapshots determines the weight. 
%While comparing, we keep track of some pointers to determine which of the modified part in the region belongs to the base (message) and which one to the result. 
%Based on these weights and pointers we statistically determine a threshold bound that decides which part of the program is altered either base or result. 
%Our confidence to define a certain change to be in the base or result depends on how much the weight is far from the threshold. 
%Larger the difference between the threshold and weights, better the confidence and determinism. 
%Thus, as the program proceeds, we determine which part of the program (either base or result) is getting modified.

%%%%%%%%%%%%%%%%%%%%%%%%%%%%%%%%%%%%%%%%%%%%%%%%%%%%%%%
%Procedure : \\

%\begin{itemize}

%\item The application should be running continuously in the background and we will be constantly monitoring the DRAM by taking snapshots through the pcileech software. We could run the victim application by giving it a set of 4 input messages/base and a key/exponent. The application will continuously work through these inputs.

%\item 

%\item Divide the 16gb DRAM space into chunks of say 100Kb (25 pages) 

%\item Run the victim application continuously.

%\item Make sure we start the parser at the point where it starts manipulating the base, result variables.

%\item Two consecutive snapshots of the same chunk will be taken and it will be compared.

%\item A heuristic will be applied for example, we note down those 5 or 6 consecutive cache lines (64 bytes per cache line) which are different. The remaining portion will be discarded.  This basically scans the entire space in small chunks to make the searching and compare more efficient and also to reduce the search space to a smaller region.

%\item Once we trim down to a smaller region we can take multiple snap shots of the region and analyze it offline.

%\item To analyze these snapshots: discard those which are similar. We would probably see one variable modified first then the other second. Try to infer the bits of the exponent.

%\end{itemize}

\section{Attack Demonstration}
\label{sec:eval}

\subsection{Experimental Setup}
% The attack has been shown on an Intel Core(TM) i7-4790 3.60GHz processor based system. BIOS version A11 for XPS 8700.
Our experiment setup uses two computer systems, one being the attacker and the other being the victim.
%\textcolor{blue}{
In our experiments, the victim system is \emph{DELL XPS 8700}, comprising of \emph{Intel Core(TM) i7-4790 3.60GHz} processor that uses \emph{Ubuntu 14.04.3 LTS} operating system with a \emph{Linux kernel 3.19.0-43-generic}, and has 16GB of main memory. 
The attacker machine is a \emph{64-bit Windows 10} based system having 8GB of main memory.
%In our experiments, the victim machine uses \emph{Ubuntu 14.04.3 LTS} operating system with a \emph{Linux kernel 3.19.0-43-generic} and 16 gigabytes of memory while, the attacker system works with \emph{64-bit, Windows 10} operating system along with 8 gigabytes memory. 
A PCI adapter module, called \emph{USB Evaluation Board}~\cite{usb}, is connected to the victim via the PCI-Express slot and acts as a compromised DMA device (cf. \figurename~\ref{fig:attack}). 
This DMA device, together with PCILeech software~\cite{pcileech}, allows the attacker to monitor victim's memory and/or take its snapshots.
To implement our attack, the PCILeech software has been extended to first find the application's address space in the victim's memory (cf. Algorithm~\ref{alg:first}), and then attack the identified address space to infer the secret key (cf. Algorithm~\ref{alg:second})). 
The above mentioned attacking algorithms run \emph{while} the victim application is executing. 
%As, mentioned in our adversarial model, we assume that either the caches in the victim system are turned off or there are \emph{write-through} caches. 
%Thus, to satisfy this assumption and to immediately propagate the updates by the application to the main memory, we use the \emph{cache flush} instruction to flush the application's data after each multiplication/squaring operation.
%\textcolor{blue}{

%To the best of our knowledge, no open source implementation of Montgomery ladder based exponentiation algorithm was found to encrypt large data files of size 128 Bytes or more.
%Hence, we used our own implementation
We have written our own C++ implementation of the Montgomery ladder based exponentiation algorithm\footnote{Available at GitHub (https://github.com/meriniamjo/RSA-Montgomery-Ladder-Implementation)} for large input sizes (128 Bytes or more) that runs on the victim system.
%of the algorithm that can be used to encrypt data files of any size.
The victim system has a BIOS version \emph{A11} which supports \emph{write-through} enabled L1 and L2 caches while disabling the L3 cache by default. 
Besides caches, any data modifications in the \emph{register file} should also be propagated to the DRAM.
The register file (usually of size 64-128 Bytes) is used by the processor to temporarily hold the operands and results during computations.
%\emph{Register spills} can be a side effect of using multiple function calls and temporary variable.
Since our implementation uses multiple temporary variables and function calls for proper execution of the algorithm, the large ``active'' working set of the application cannot fit into the register file and results in \emph{register spills}.
%Due to large input data and modifications done to temporary variables, registers would not be able to hold such large data. 
%Hence, resulting in \emph{register spills}, which supports our attacking methodology by allowing data to be propagated to the DRAM.
Hence, any updates made by the application are immediately propagated -- through register file and caches -- down to the main memory as each multiplication/squaring write operation in performed.
%}
Section~\ref{sec:results} explains the step by step details about how the attack is launched.

\subsection{Experimental Results}
\label{sec:results}

In order to take memory snapshots via PCI module and the PCILeech software, 
%infer the secret key bits, the attacker needs to read the whole memory. 
%To accomplish this task, 
the attacker first needs to load a kernel module into the victim system via the PCI module itself. 
Notice that the attacker does not require any extra privileges to do so. 
We use the following command via PCILeech software to load the kernel into victim's DRAM. 
When the kernel is loaded, an address is spitted out by the software, which shows where the module resides in the victim's memory. 
Loading the kernel into memory is a rapid process and takes only a few milliseconds to complete the process. 

\begin{tcolorbox}
\small{\verb|D:\>|}\texttt{pcileech kmdload $-$kmd LINUX\_X64}\\
\texttt{\small {KMD: Code inserted into the kernel}}\\
\texttt{\small {KMD: Execution received - continuing ...}}\\
\texttt{\small {KMD: Successfully loaded at 0x1b54a000}}\\
\small{\verb|D:\>_|}
\end{tcolorbox}

In the meantime, the Montgomery's ladder exponentiation algorithm is run on the victim machine using a 128 byte (1024 bits) message along with a secret key of 64 bytes (512 bits). 

\begin{tcolorbox}
\texttt{\small {[user@victim]\$ ./montgomery\_exponentiation}}
\end{tcolorbox} 

With the application running and the kernel module loaded into victim's memory, we proceed to find the potential regions in the DRAM which are being accessed frequently by taking multiple snapshots. 
To retrieve these snapshots, we issue the \emph{pagefind} command shown below which uses the loaded kernel module's address to access the victim's full memory.

We integrated the \emph{pagefind} command into the PCILeech software to iteratively find regions getting modified persistently. 
\emph{pagefind} narrows down the selected regions to a single page by constantly monitoring and comparing the changes being made, and returns the address of the page where application's array data structures are defined. 
This step corresponds to \emph{Application's Address Space Identification} phase of the attack (cf. Algorithm~\ref{alg:first}) and is the most time consuming phase. 
To read the whole memory, comparing their respective snapshots and narrowing down to a single page of 4KB from 16GB search space takes $\sim$3 minutes and 30 seconds.
%\texttt{\small {D:$\setminus$pcileech$>$pcileech pagedisplay} \scriptsize{$-$min} \small{0xd271c000}}\\\\
\begin{tcolorbox}
\small{\verb|D:\>|}\texttt{\small {pcileech pagefind $-$kmd 0x1b542000}}\\
\texttt{\small {Matching Pattern ...}}\\
\texttt{\small {Page Finding: Successful.}}\\
\texttt{\small {Total\_Time $=$ 210199 Milliseconds}}\\
\texttt{\small {Victim Page Address : 0xd271c000}}\\
\small{\verb|D:\>_|}
\end{tcolorbox}

As shown above, from the first phase we retrieve the address of the page where application's data structures are stored. 
Proceeding towards our second and third step namely \emph{Distinguishing Local Variables} and \emph{Inferring the Secret Key} (cf. Section \ref{sec:step2}, \ref{sec:step3}), we use another integrated command \emph{pageattack}.
It first takes a predefined number of snapshots of the application page provided by the first step, and distinguishes the message ($R_1$) and algorithm result ($R_0$) from the rest of the stale data, residing on the memory page. 
It then uses the order of changes in $R_0$ and $R_1$ to infer the secret key.

\begin{tcolorbox}
\small{\verb|D:\>|}\texttt{\small {pcileech pageattack $-$min 0xd271c000}}\\
\texttt{\small {Attack Successful.}}\\
\texttt{\small {Total\_Time $=$ 3596 Milliseconds}}\\
\texttt{\small {Inferred Key is:}}\\
\texttt{\small {1a 4b 28 41 e6 27 d4 7d}}\\
\texttt{\small {72 c3 40 79 be 1f 6c 35}}\\
\texttt{\small {ca 3b 58 b1 96 17 04 ed}}\\
\texttt{\small {22 b3 70 e9 6e 0f 9c a5}}\\
\texttt{\small {7a 2b 88 21 46 07 34 5d}}\\
\texttt{\small {d2 a3 a0 59 1e ff cc 15}}\\
\texttt{\small {2a 1b b8 91 f6 f7 64 cd}}\\
\texttt{\small {82 93 d0 c9 ce ef fc 85}}\\
\small{\verb|D:\>_|}
\end{tcolorbox}

This final step takes $\sim$3.6 seconds to complete and returns the complete 512 bit secret key learned from only the write access patterns. 
Combing the times associated with all the attack phases, the total attack time comes out to be $\sim$3 minutes, 34 seconds. 

%To distinguish between $R_0$ and $R_1$, we use the page histogram technique (mentioned earlier) and assign weights to the modifying snapshots and determine a threshold based on it. If the snapshot difference results in a weight greater than the threshold, we consider that modification as a change happening in $R_0$ otherwise, $R_1$ (as shown in algorithm~\ref{alg:second}). Subsequently, we align them in a pair format and based on their transitioning i.e. $R_0$ $\rightarrow$ $R_1$ or $R_1$ $\rightarrow$ $R_0$, assign the key bit value either 1 or 0, respectively. 

\section{Leakage under Caching Effects}\label{sec:cache_attack}

%\subsection{Potential Attacks with Caching Effects}

%1)What happened without caching effects and what will you discuss regarding the caching effects.
%2)Talk about a potential attack where cache striding is used and how contention is created over one set.
%3)How Gaussian eliminations comes to this cache striding example attack.
%4)What information can be leaked when you guess which matrix is being used 
%5)what about a large image pixel matrix where you would need a smaller strided access that hits the same set.

In view of our proposed attack on Montgomery ladder based exponentiation algorithm, the updates to the application data should always be available in the DRAM of the victim system before an attacker issues a memory snapshot request.
%The attack over Montgomery ladder based exponentiation algorithm proposed in this paper works as expected if it runs 
This is only possible if the victim system has \emph{write-through} enabled cache hierarchy or the caches are disabled altogether.
Whereas, on the other hand, modern processors typically consist of large on-chip \emph{write-back} caches where the updates to application's data are only visible in DRAM once the data is evicted from the last level cache (LLC).
Thus the attack proposed in Section~\ref{sec:attack}, the caching effects are not catered for, which introduces `noise' to the precise write-access sequence inferred earlier, hence making the attacker's job difficult.
A possible workaround to deal with such caching effects is to collect several `noisy' sequences of memory snapshots and then run correlation analysis on them to learn the precise write-access pattern.
Furthermore, if the adversary is also a user of the same computer system, it can flush the system caches frequently to reduce the noise in write-access sequence even further.

%In a nutshell, the proposed attack works efficiently when each modification is propagated to the backing storage (DRAM) right away as the write is performed. 
%However, if the system does not have \emph{write-through} enabled cache policy, the inferred key bits may be wrong as the data may never get committed to the main memory.
%In order to leak privacy under the caching effects, a different attack strategy is required. 
%One such 
Another (more efficient) attack scenario under write-back caches would be when the application has a strided memory access pattern that causes contention over the cache sets, and hence forces its own data to be evicted to make room for the new data in the cache. 
In the following subsection, we discuss how such a strided memory access pattern can lead to evictions to the DRAM which could potentially leak private information.

\subsection{Memory Striding and Cache Set Contention}\label{sec:striding}

% ------ explain general strided example in terms of array or matrix----------------

\begin{figure}[!t]
\centering
\subfloat[Accessing an array with a stride of 4.]{\includegraphics[width=1.2in]{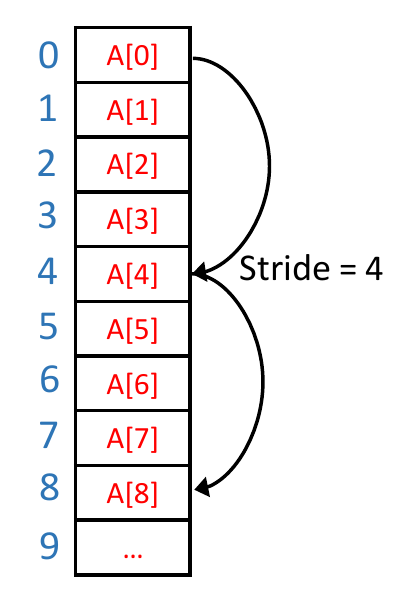}
\label{fig:stride_b}}
\hfill
\subfloat[A 2-way set-associative cache.]{\includegraphics[width=2in]{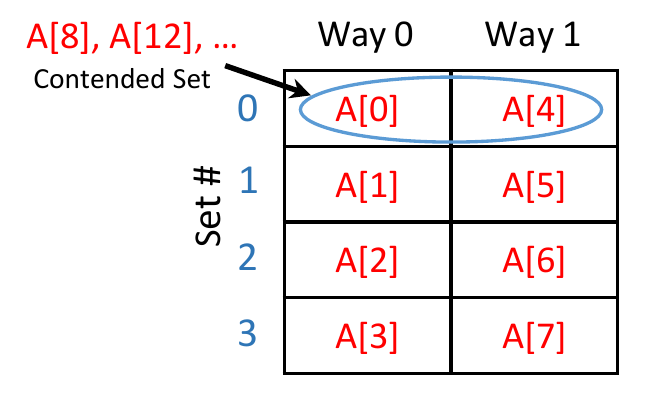}
\label{fig:stride_a}}
\caption{A strided memory access pattern causing contention on a single cache set.}
\label{fig:cache}
\end{figure}

A strided memory access pattern is the one where each request to the memory is for the same number of bytes, and the access pointer is incremented by the same amount between each request. An array accessed with a stride of exactly the same size as the size of each of its elements results in accessing contiguous locations in the memory. 
Such access patterns are said to have a stride value of 1. 
\figurename~\ref{fig:stride_b} shows a \emph{non-unit} striding access pattern in which the elements $0, 4, 8, 12, \cdots$ of an array $A$ are accessed.
This access pattern has a stride value of $4$. 

% ------ Link it with caches ------------------------------------------

%The total cache capacity is divided among cache lines of equal size in bytes. 
%Now, the length of the memory access stride, relative to a cache line size can affect the location where the cache is accessed. 
%For a stride much smaller than a cache line size, many consecutive accesses could go to the same cache line per set. 
%If the stride is larger than a cache line size, the successive accesses could go to different cache lines therefore, causing accesses to fall into the same set of the cache. 
%The following section demonstrates such a toy example.

%\subsection{Toy-Example}\label{sec:example}

%\begin{figure}[t!]
%\centering
%\includegraphics[width=0.9\columnwidth]{figs/cache.pdf}
%\caption{A 2-way set associative cache mapped to a strided memory access pattern.}
%\label{fig:cache}
%\end{figure}

Consider a simple system which has a 2-way set-associative \emph{write-back} cache with a total capacity of 8 cache lines, as shown in  \figurename~\ref{fig:stride_a}.
The strided access pattern from \figurename~\ref{fig:stride_b} accesses every $4i_{th}$ element of the array $A$, where $i=0,1,2,\cdots$.
Assuming that each element of $A$ is of size equal to the cache line size, for a simple \emph{modulus based} cache hash function, the elements $A[0], A[4], A[8], \cdots$ are mapped to the same set causing contention over \emph{set $0$}.
Since the cache associativity is only 2, this access sequence causes evictions from the cache when both ways of set $0$ contain valid cache lines.
Similarly, elements $A[1], A[5], A[9], \cdots$ map to \emph{set $1$}, and this access sequence will cause evictions from set 1, and so on.
In other words, such write-access sequences are still propagated almost immediately to the next level in the memory hierarchy (e.g., DRAM) even under write-back caches, which could potentially leak information.
This is an artifact of the cache implementation combined with the striding access pattern of the application.

It must be noted that, not all evictions result in updates to the main memory.
Typically, only \emph{dirty} cache lines, caused by data writes, evicted from the cache are propagated to the main memory.
\emph{Clean} evictions from the cache are simply discarded resulting in no change in the main memory since it already contains a clean copy of the data.
%The same eviction process applies to the other sets as well, depending on the stride length. 

Assume that an application generates two distinguishable striding write-access patterns that result in contention at two different cache sets, leading to evictions from the cache.
Consequently, the resulting write-access access sequence will be revealed to an adversary who is capable of monitoring changes in the main memory, potentially resulting in privacy leakage.
%As the application is public, the adversary could observe the contended sets and infer critical information via reverse engineering.
%That critical information could be the state of the application or the data inferred from the access patterns.
%The adversary could reverse engineer the secret information by observing the set from which evictions occur to the secret. 
%Consider an example where an adversary observes the evictions from set 0 and 1 while the cache lines in other sets remain as they were. Such a scenario could lead to two distinguishable access patterns of an application hence, inferring secret information about it. 
%Since the application itself is public, the adversary could infer some information about the application state or data from the two patterns.

\subsection{Striding Application: Gaussian Elimination}

In Section~\ref{sec:striding} we discussed, using a toy example, how a strided access pattern can lead to information leakage.
Now we present a realistic example which has such a striding access pattern, and later in Section~\ref{sec:mceliece} we show how such a pattern can be exploited to learn private information.
%To support the claim that strided access patterns can potentially leak information, 
We consider the application of \emph{Gaussian Elimination} of large binary matrices carrying substantial amount of information. 
Clearly, these large matrices cannot fit into the caches, therefore there will be cache evictions as a result of Gaussian elimination operations. 
%Hence, causing contention over the cache sets.
%In order to exploit the striding pattern mentioned in section \ref{sec:example}, we will utilize matrices that stores large amount of information that will not fit fully into the caches. One such matrix operation that also leads to changes in the matrix is Gaussian elimination.
%The following subsection discusses the algorithm of Gaussian elimination, subsequently we show an attack over a binary matrix used for the encryption and decryption of data.

%can be then we will talk about how these operations could lead to potential secret information leakage.

%In the following section, we discuss applications which use strided memory access patterns leading to potential secret information leakage.

%\subsubsection{Algorithm}

%\textcolor{blue}{In order to exploit the striding pattern mentioned in section \ref{sec:example} we will utilize matrices that stores large amount of information that will not fit fully into the caches. One such matrix operation that also leads to changes in the matrix is Gaussian elimination.}
%(cf. Algorithm~\ref{alg:third})
Gaussian elimination a.k.a. row reduction is a method for solving system of linear equations by the use of matrices in the form $Ax = B$. 
Row reduction is done by doing a series of elementary row operations which modify the matrix until it forms an upper triangular matrix, i.e., elements underneath the main diagonal are zeros. 
Different types of elementary row operations include swapping two rows, multiplying a row by a non-zero number and adding a multiple of one row to another. 
The upper triangular matrix formed out of these operations will be in row echelon form. 
When the leading coefficient (pivot) in each row is 1, and every column containing the leading coefficient has zeros elsewhere, the matrix is said to be in reduced row echelon form.
The Gaussian elimination algorithm consists of two processes, one being \emph{forward elimination} that converts the matrix to row echelon form and the other is \emph{backward substitution} that calculates values of the unknowns. These processes result in solving the linear equation.

Gauss-Jordan elimination uses a similar approach for finding the inverse of a matrix.
For a $n \times n$ square matrix $S$, elementary row operations can be applied to reduce the matrix into reduced echelon form, and furthermore, for computing the matrix inverse if it exits.
Initially, the $n \times n$ identity matrix $I$ is augmented to the right of $S$, forming a $n \times 2n$ block matrix $[S | I]$.
Now, upon applying the row operations, the left block can be reduced to the identity matrix $I$ if $S$ is invertible. 
This gives $S^{-1}$ which is the right block of the final matrix. 
In a nutshell, we continue performing row operations until $[S | I]$ becomes $[I | S^{-1}]$.

Consider that the matrix under elimination is stored in a \emph{column-contiguous} manner in the computer system's main memory.
In other words, each column occupies a contiguous chunk of memory equal to the column size, after which the next column resides, and so on.
When consecutive elements of a row of this matrix are accessed during a row operation, the corresponding memory access pattern results in a striding sequence, where the stride length is equal to the column size.
If the stride length is such that it creates contention on particular cache sets corresponding to particular rows, this would reveal the modified row, which in turn could potentially leak the binary matrix itself (cf. Section~\ref{sec:mceliece}).

%\begin{algorithm}[!t]
%\caption{Gaussian Elimination to Solve $Ax = b$} \label{alg:third}
%\textbf{Inputs:} $A$: $n$ X $n$ matrix, $b$: $n$-vector. \\
%\textbf{Output:} $x$: Output solution vector . \\
%\textbf{Start:}\\
%\textbf{Forward Elimination}
%\begin{algorithmic} [1]
%\For{$k = 1$ to $n-1$} %\Comment{Loop over all rows in matrix except last.}
%\For{$i = k+1$ to $n$} %\Comment{Loop over all rows below diagonal position(k,k).}
%\State $m_{ik} = a_{ik}/a_{kk}$ %\Comment{Compute multiplier for row (i) column (k)}
%\For{$j = k+1$ to $n$} %\Comment{Loop over all columns to the right of the diagonal position(k,k).}
%\State $a_{ij} = a_{ij} - m_{ik}*a_{kj}$ %\Comment{Update coefficient for row (i) column (j)}
%\EndFor
%\State $b_i = b_i - m_{ik}*b_k$ %\Comment{Update the right hand side for row (i) column (j)}
%\EndFor
%\EndFor
%\end{algorithmic} 
%\textbf{Backward Substitution}
%\begin{algorithmic} [1]
%\State $x_{n} = b_{n}/a_{nn}$ %\Comment{Compute last unknown}
%\For{$i = n-1$ to $1$} %\Comment{Loop backwards over all rows except last}
%\State $sum = 0$
%\For{$j = i+1$ to $n$} %\Comment{Loop over all columns to the right of the current row}
%\State $sum = sum + a_{ij} * x_{j}$
%\EndFor
%\State $x_i = (b_i - sum)/a_{ji}$
%\EndFor
%\end{algorithmic} 
%\end{algorithm}

%---- explain how the attack works---------

\subsection{Attacking McEliece Public-Key Cryptosystem}\label{sec:mceliece}

\begin{figure}[!t]
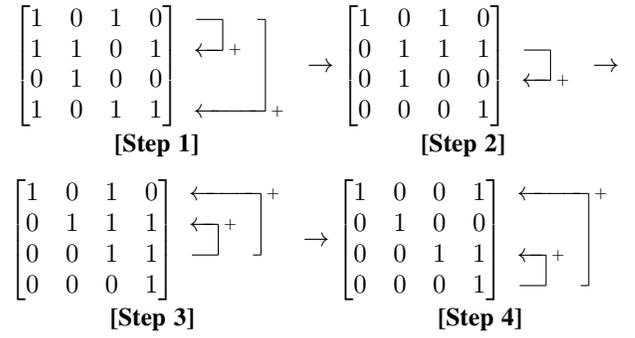

\centering
\[
\begin{sbmatrix}{\textbf{\lbrack Step 1\rbrack}}
    1       & 0 & 1  & 0 \\
    1       & 1 & 0  & 1 \\
    0       & 1 & 0  & 0 \\
    1       & 0 & 1  & 1
\rowops
\add{0}{1}
\add{0}{3}
\end{sbmatrix}
\rightarrow
\begin{sbmatrix}{\textbf{\lbrack Step 2\rbrack}}
    1       & 0 & 1  & 0 \\
    0       & 1 & 1  & 1 \\
    0       & 1 & 0  & 0 \\
    0       & 0 & 0  & 1
\rowops
\add{1}{2}
\end{sbmatrix}
\rightarrow
\]
\[
\begin{sbmatrix}{\textbf{\lbrack Step 3\rbrack}}
    1       & 0 & 1  & 0 \\
    0       & 1 & 1  & 1 \\
    0       & 0 & 1  & 1 \\
    0       & 0 & 0  & 1
\rowops
\add{2}{1}
\add{2}{0}
\end{sbmatrix}
\rightarrow
\begin{sbmatrix}{\textbf{\lbrack Step 4\rbrack}}
    1       & 0 & 0  & 1 \\
    0       & 1 & 0  & 0 \\
    0       & 0 & 1  & 1 \\
    0       & 0 & 0  & 1
\rowops
\add{3}{2}
\add{3}{0}
\end{sbmatrix}
\]
\caption{The Gaussian Elimination process on a $4\times4$ binary matrix.}
\label{fig:elimination}
\end{figure}

%We consider a large matrix such that the coefficients are stored contiguously in an array, column wise.
%The elements in the column fill the cache such that starting index of each column hits the same cache set, say set 0.
%Hence, any changes to the first row of the matrix could lead to evictions from the same cache set 0. 
%Seeing from which set we see writes in the DRAM we can infer which row of the matrix is getting modified.
%Once we have a trace of the set of row modifications we could reverse engineer the operations to leak some portion of vector $x$.

McEliece public key cryptosystem~\cite{secretInverseMatrix}, an asymmetric encryption algorithm, uses an error correcting code for a description of the private key. 
%McEliece public key cryptosystem~\cite{secretInverseMatrix} uses a linear error-correcting code to encrypt/decrypt the data, which has a fast decoding algorithm, namely a Goppa code. 
This encryption uses a fast and efficient decoding algorithm, namely a Goppa code and hides the structure of the code by transformation of the generator matrix. This transformation yields the public key and the structure of the Goppa code together with the transformation parameters, which further provides the trapdoor information. 
For a linear code \textit{C}, generator matrix $G$, random invertible matrix $S$ and random permutation matrix $P$, the matrix $G^\ast = SGP$ is made public while $P$, $G$, and $S$ form the private key. 
A message $m$ is encrypted along with a random error vector using the equation $c = mG^\ast + e$, where $c$ refers to the ciphertext. 
In the decryption process, we compute $c^\ast = cP^{-1}$, decode $c^\ast$ to $m^\ast$ by the decoding algorithm, and lastly compute $m = m^\ast S^{-1}$. 
Notice, that $S$ is a private binary matrix whose inverse is used to recover the message $m$. 
Any system carrying out this encryption/decryption process could either store the matrix inverse (for better performance) or calculate the inverse during the run time. However, the latter could lead to the leakage of the binary matrix via write-access patterns during the inverse computation.
In this section we will demonstrate how performing Gauss-Jordan elimination \cite{binaryGaussElimin} on the binary secret matrix $S$ could lead to its complete exposure as a consequence of cache striding and cache set contention as shown in section \ref{sec:striding}.

%As shown in \cite{secretInverseMatrix} inverse of a secret matrix can be used for encryption and decryption of data. 

For the ease of demonstration we consider a $4 \times 4$ binary matrix. 
The elements stored in the main memory are column contiguous.
We assume that each element of the matrix is \emph{cache line aligned} for performance reasons.
In other words, each element is stored in a unique cache line in order to avoid false sharing within a cache line.
%and the elements stored in memory in a column contiguous fashion.
%\textcolor{red}{This allows each row of the matrix to map to a single set of a 4 set 2-way set associative cache. 
%Since there are 4 elements in a row and only 2 ways in the set, accessing the whole row would eventually lead to evictions from the set.}
Considering a system with a 2-way 4-set associative cache, each row of the matrix is mapped to one cache set due to the cache structure (cf. \figurename~\ref{fig:stride_a}).
%Our approach towards the attack assumes that, from the write access patterns, the adversary has the capability to infer the cache set of a 2-way 4-set associative cache from which evictions happen. Therefore, mapping the set to the row of the matrix.
The elimination process to obtain the inverse of the binary matrix $S_{4 \times 4}$ is shown step by step in \figurename~\ref{fig:elimination}. 
After these row operations, we obtain an identity matrix $I_{4 \times 4}$.

%\rightarrow
%$I_{4 \times 4}$

%\begin{spmatrix}{\lbrack 5 \rbrack}
%    1       & 0 & 0  & 0 \\
%    0       & 1 & 0  & 0 \\
%    0       & 0 & 1  & 0 \\
%    0       & 0 & 0  & 1
%\end{spmatrix}

In each of the above 4 steps, the corresponding pivot row is added to another row or rows.
For example, in $\textbf{Step \lbrack1\rbrack}$ row $1$ is added to row $2$ and row $4$. Similarly, row $2$ is added to row $3$ in $\textbf{Step \lbrack2\rbrack}$. 
As a row operation is performed, the elements of the target row are modified and result in cache line evictions, since accessing a whole row causes contention over the corresponding set it is mapped to, and a cache set can only store 2 elements of a row at a time.
For instance, in $\textbf{Step \lbrack1\rbrack}$, row $2$ and row $4$ are modified causing contention and evictions from set $1$ and $3$ respectively.
Consequently, an adversary can learn the identifier of the row being updated during each row operation by monitoring the address space in which any updates take place, and then linking it back to the row number.
Now, by definition of the elimination algorithm, all column elements corresponding to rows that undergo addition operations can be inferred as $1$s, and the remaining ones as $0$s. 
Hence, in each of $\textbf{Step \lbrack1\rbrack}$, $\textbf{Step \lbrack2\rbrack}$, $\textbf{Step \lbrack3\rbrack}$ and $\textbf{Step \lbrack4\rbrack}$, we infer the corresponding pivot column to be $ C_1 = \lbrace1,1,0,1\rbrace$, $C_2 = \lbrace0,1,1,0\rbrace$, $C_3 = \lbrace1,1,1,0\rbrace$ and $C_4 = \lbrace1,0,1,1\rbrace$ respectively.

\begin{figure}[!t]
\centering
\[
C_2= 
\begin{sbmatrix}{}
    0  \\
    1  \\
    1  \\
    0  
\rowops
\add{0}{1}
\add{0}{3}
\end{sbmatrix}
\rightarrow
\begin{sbmatrix}{}
    0  \\
    1  \\
    1  \\
    0  
\end{sbmatrix}
\textbf{$= S_2$}
\] 
\[
C_3 = 
\begin{sbmatrix}{}
    1  \\
    1  \\
    1  \\
    0  
\rowops
\add{1}{2}
\end{sbmatrix}
%\rightarrow
\begin{sbmatrix}{}
    1  \\
    1  \\
    0  \\
    0 
\rowops
\add{0}{1}
\add{0}{3} 
\end{sbmatrix}
\rightarrow
\begin{sbmatrix}{}
    1  \\
    0  \\
    0  \\
    1 
\end{sbmatrix}
\textbf{$= S_3$}
\] 
\[
C_4 =  
\begin{sbmatrix}{}
    1  \\
    0  \\
    1  \\
    1  
\rowops
\hspace{-6pt}
\add{2}{1}
\add{2}{0}
\end{sbmatrix}
\hspace{-8pt}
%\rightarrow
\begin{sbmatrix}{}
    0  \\
    1  \\
    1  \\
    1  
\rowops
\hspace{-6pt}
\add{1}{2}
\end{sbmatrix}
\hspace{-8pt}
%\rightarrow
\begin{sbmatrix}{}
    0  \\
    1  \\
    0  \\
    1 
\rowops
\hspace{-6pt}
\add{0}{1}
\add{0}{3} 
\end{sbmatrix}
\hspace{-10pt}
\rightarrow
\begin{sbmatrix}{}
    0  \\
    1  \\
    0  \\
    1 
\end{sbmatrix}
\hspace{-8pt}
\textbf{$= S_4$}
\] 
\caption{Back Substitution process to recover secret binary matrix $S$.}
% $S^{\ast}$.}
\label{fig:back_subs}
\end{figure}

Notice that $C_2$, $C_3$, and $C_4$ obtained in the above steps show the respective intermediate forms of the corresponding columns of $S$ during the elimination process. 
These values, however, can be used to recover the original column values of matrix $S$ through \emph{back substitution} process, as shown in \figurename~\ref{fig:back_subs}.
In this process, each column $C_i$ undergoes the row operations performed (and inferred) in each of the steps $\textbf{Step \lbrack i-1\rbrack}, \textbf{Step \lbrack i-2\rbrack}, \cdots, \textbf{Step \lbrack1\rbrack}$, precisely in this order.
For example, $C_2$ undergoes addition of row $1$ to rows $2$ and $4$, while $C_3$ performs addition of row $2$ to row $3$ along with the addition of row $1$ to rows $2$ and $4$. 
Upon completion of the back substitution process, the complete secret matrix $S$ is recovered by the adversary.

%In a practical seting, a larger matrix will be used in the crypto system as a secret matrix to encrypt or decrypt data blocks.
%Also, the cache will be relatively larger. 
%Suppose, we have a $512$ KBytes $8$-way set associative cache with a cache line size of $64$ bytes and a large $1024 \times 1024$ secret matrix whose inverse needs to be computed.
%Assuming each element of the binary matrix to be one byte, $64$ elements of a row will fit into one cache line. 
%Hence, $16$ such cache lines will form one row of the $1024 \times 1024$ matrix. 
%For such a cache configuration, there are $1024$ sets which will map to the $1024$ rows of the matrix. 
%Notice, if the cache lines forming the matrix are column contiguously stored in the memory, each cache line corresponding to a row will take a stride of $64$ KBytes. Since there are only $8$ ways in the set, making changes to $16$ cache lines for a row corresponding to a set will eventually lead to evictions from that set. Consequently, contending the cache set and the adversary can infer the secret matrix in a similar fashion as done for the $4 \times 4$ matrix.

\section{Discussion}\label{sec:discuss}

\subsection{Potential Threats in Database Applications}
%applications for linked list and binary tree}

Database applications are ubiquitously being used to facilitate simultaneous updates and queries from multiple users. 
%Moreover, there has been an explosion of new systems for data storage and management ``in the cloud''. 
Such databases and cloud servers could store high volumes of data regarding an organization's operations. 
%They typically process relatively high volumes of updates. 
Examples include databases that record contact and credit information etc. about employees in an organization. 
In such databases, the \textit{key} or the \emph{record locater} of a piece of information, for instance, can be the social security number of the employee.
%Similarly, there could be database servers that hold information about employees or keep track of the organization's money, accounting and financial dealings. Such databases where all the employee records are stored, the key of the record can be the social security number (SSN) associated with data such as name, address, salary, benefits etc. 
%Another application of database servers can be for a medical institution which keeps track of patient's information where the key can be the SSN again, and the data being patient's history. 
Leaking the SSN of an employee from above mentioned databases can be detrimental~\cite{news}. 
%Consider an example, where the attacker could be an account holder in the organization and works alongside another attacker to invade the system and leak some confidential information. 
%This can be potentially done when the data is arranged in the following ways; 
%We consider the data to be arranged in the following ways, which can potentially leak the private information;
% cloud setting where all the users can create an account or log in and can attack the system. 
In the following, we discuss how write-access patterns could potentially leak private information stored in two commonly used data structures.

\subsubsection{Linked List} \label{sec:linked}
Linked list data structure is a collection of a group of data elements, called nodes, which together represent a sequence.
%It allows insertion and removal of nodes at any point in the list, and can do so with a constant number of operations. 
%if the link previous to the link being added or removed is maintained during list traversal.
Consider the users' encrypted private information stored in a singly linked list based on the increasing order of their social security numbers (SSN).
The linked list is stored in the victim system's disk.
The adversary's goal is to collect as many valid SSNs (which represent identities of real humans) from this linked list as possible.

Assume that two attackers are working in tandem and both of them are users of the linked list database mentioned above.
Since for each search/update operation, the linked list is traversed started from the first node up to the node being searched/updated, all these nodes will be loaded from the disk into the DRAM in the specific ascending order.
Meanwhile, since the adversaries can monitor the DRAM snapshots throughout this process, they learn the exact order of each node in the linked list. 
Now, the adversaries can insert/update their own information in the database while monitoring the resulting memory write pattern.
This allows them to link their own SSNs to two particular nodes in the linked list, separated by potentially a small number of other nodes.
Crucially, this gives the adversaries a potentially small range of valid SSNs bounded by the two adversarial SSNs.
Next time, whenever a user's information is updated which results in modifying a node between two adversarial nodes, the adversaries can brute force the small range of SSNs to find the valid SSN of the particular user.

\subsubsection{Binary Search Tree (BST)} 

Binary search tree (BST) data structure stores data in memory while allowing fast lookup, addition and removal of the stored data. 
%It can be used to implement lookup tables that allow finding an item by its sorted key.
To perform a lookup/update operation, BST looks for a key by traversing the tree from root to leaf while choosing left or right child at each level based on the key to be searched.

Consider a similar scenario as in Section~\ref{sec:linked} where the data is stored in a BST instead, and assume that the attackers know the initial layout of the BST in the victim system's memory.
Since traversing the tree involves only read accesses, this won't leak any information in our model.
However, in case of write accesses (inserting or updating a node), the position where the node is inserted or updated will be leaked.
Following a similar strategy as Section~\ref{sec:linked}, the adversary can deduce a range of SSNs by inserting his own nodes in the BST.
A larger group of individuals acting as adversary can deduce further smaller ranges of SSNs resulting in easier attack.
 
Notice that during this type of attack, the subtree corresponding to the range of SSNs found by the adversary should not be evicted from DRAM.
If, for some reason, the pages corresponding to the vulnerable subtree are swapped out to disk by the OS, the layout of the subtree in the DRAM could be different once it is loaded in DRAM next time.
In this case, the adversary will need to relearn the BST layout in DRAM by monitoring several updates over a long period of time.
%Also notice that, it is necessary for all this process to happen before the pages are removed from DRAM because reading the files from disk to DRAM again will not show the same structure of the tree in memory and the attackers will not be able to know whether a node is placed in the left or the right subtree. Such a situations make this attack non-trivial.

%Moreover, as this process does not involve any write accesses thus

%the attack is not trivial. As,
%In case of a binary search tree first let’s assume the attacker knows the layout of the tree in DRAM. When a node is updated the tree is traversed till the desired node is found. In this process there will only be reads from the tree and no writes for us to monitor. So we will not be leaking much helpful information. 

%To know the layout of the tree in DRAM the attacker needs to read the memory right from the time the tree is formed. This is not very usual. . The data will be encrypted so its difficult to know the range of SSN value which needs to be leaked. The best we could do is insert two nodes with close range and wait for some insertion in between these nodes. This is assuming the attacker knows the layout of the tree in DRAM.

\subsection{Future Work: Countermeasures for Our Attack}
One approach to prevent privacy leakage via write-access pattern leveraging DMA based attacks, as demonstrated in this paper, could be to block certain DMA accesses through modifications in the DRAM controller.
However, this approach poses complexity in terms of how to determine which accesses to allow and which ones to block. 
Furthermore, it requires this `extended' DRAM controller to be included in the trusted computing base (TCB) of the system which is undesirable.

Another strong candidate is Oblivious RAM which is a well known technique to prevent privacy leakage via memory access patterns.
Although, current so called \emph{fully functional} ORAMs, which obfuscate both read and write patterns, offer a possible countermeasure (at a cost of performance penalty) against the attack we demonstrated. 
However, the extra protection (read pattern obfuscation) offered by these approaches is an overkill for current attack scenario and incurs redundant performance penalties.

%\textcolor{blue}{
A better alternative could be a \emph{write-only} ORAM~\cite{hive, li2013write, haider2016flat} which only obfuscates write-access patterns and not the reads. 
This technique offers far better performance than a fully functional ORAM under such weaker adversarial models. It has been shown that \emph{write-only} ORAM has an optimal asymptotic communication overhead of $O(1)$ as compared to the fully functional ORAM schemes, which are asymptotically $\Omega(\log{}n)$~\cite{hive}.
%}.

\section{Conclusion}

Privacy leakage via purely write-access patterns is less obvious and not extensively studied in the current literature.
We demonstrate a real attack on Montgomery's ladder based modular exponentiation algorithm and infer the secret exponent by just learning the write access patterns of the algorithm to the main memory.
We adapt the traditional DMA based exploits to learn the application's write access pattern in a reasonable time.
Our attack takes just 3 minutes and 34 seconds to learn 512 secret bits from a typical Linux based victim system.
A possible attack on McEliece public-key cryptosystem has also been presented.
We discuss some possible countermeasures to prevent such attacks.
Further research towards developing efficient countermeasures is left as future work.

\label{sec:conclusion}

%*******************************************************%

%% if have a single appendix:
%%\appendix[Proof of the Zonklar Equations]
%% or
%%\appendix  % for no appendix heading
%% do not use \section anymore after \appendix, only \section*
%% is possibly needed
%
%% use appendices with more than one appendix
%% then use \section to start each appendix
%% you must declare a \section before using any
%% \subsection or using \label (\appendices by itself
%% starts a section numbered zero.)
%%
%
%
%\appendices
%% you can choose not to have a title for an appendix
%% if you want by leaving the argument blank
%\section{}

\ifCLASSOPTIONcompsoc
  \section*{Acknowledgments}
\else
  \section*{Acknowledgment}
\fi

This work was partially supported by NSF grant CNS-1413996 for MACS: A Modular Approach to Cloud Security.

\ifCLASSOPTIONcaptionsoff
  \newpage
\fi


\bibliographystyle{IEEEtran}
\bibliography{bib/dma_bib} 


\begin{IEEEbiography}[{\includegraphics[width=1in,height=1.25in,clip,keepaspectratio]{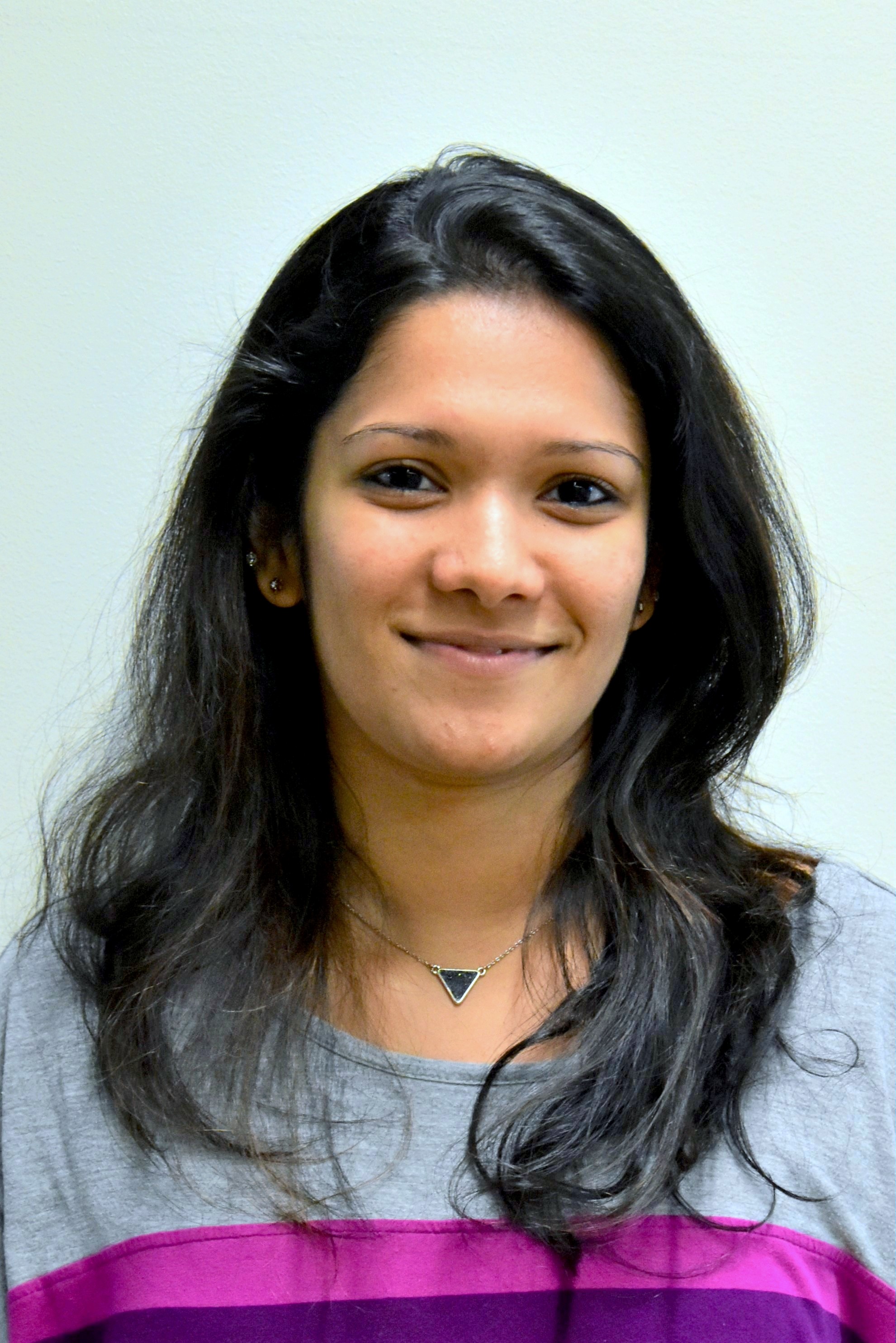}}]{Tara Merin John}
is a M.S. student in Electrical \& Computer Engineering at University of Connecticut.	
Her research interests include Computer Architecture, Hardware Security and Embedded Systems Design. 
She did her Bachelors in Electronics and Communication Engineering from University of Kerala, Kerala, India. 
\end{IEEEbiography}
\vspace{-10mm}
\begin{IEEEbiography}[{\includegraphics[width=1in,height=1.25in,clip,keepaspectratio]{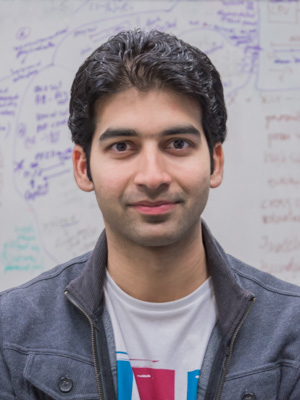}}]{Syed Kamran Haider}
is a Ph.D. student in Electrical \& Computer Engineering at University of Connecticut.	
His research interests include Computer Architecture and Hardware Security. 
Before coming to UConn, he did his Masters in Embedded Computing Systems as a joint degree from NTNU Trondheim, Norway and TU Kaiserslautern, Germany. 
\end{IEEEbiography}
\vspace{-10mm}
\begin{IEEEbiography}[{\includegraphics[width=1in,height=1.25in,clip,keepaspectratio]{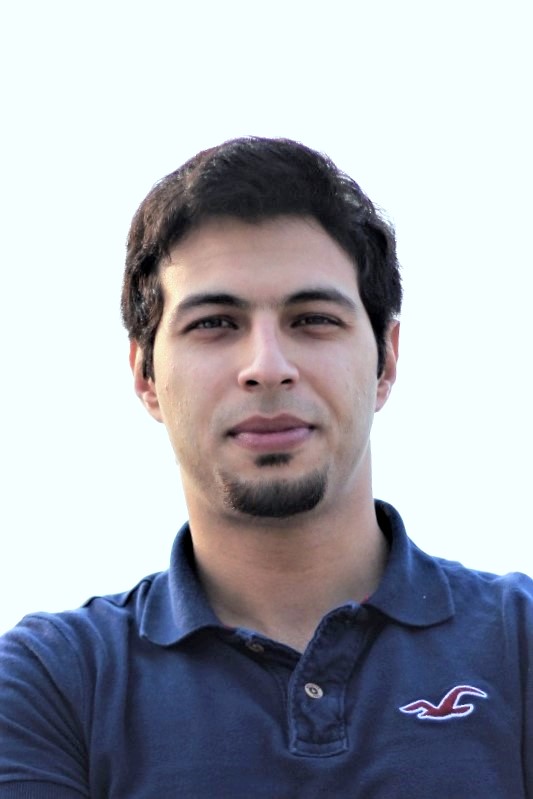}}]{Hamza Omar}
is a Ph.D. student in Electrical \& Computer Engineering at University of Connecticut.	
His research interests include Computer Architecture, Hardware Security and Approximate Computing. 
Prior to joining UConn as a graduate student, he did his Bachelors in Electrical Engineering from University of Engineering \& Technology, Lahore, Pakistan.
\end{IEEEbiography}
\vspace{-10mm}
%\newpage

\begin{IEEEbiography}[{\includegraphics[width=1in,height=1.25in,clip,keepaspectratio]{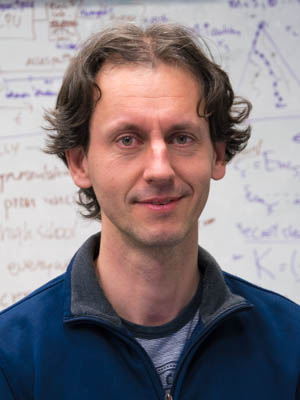}}]{Marten van Dijk}
is a Charles H. Knapp Associate Professor of Electrical and Computer Engineering at the University of Connecticut. 
Prior to joining UConn, he was a Research Scientist at MIT, RSA Labs, and Phillips Research.
He has been awarded 2015 ACM/IEEE A. Richard Newton Technical Impact award in Electronics Design Automation.
He has won several best paper awards.
\end{IEEEbiography}


%\vfill


\end{document}